# Carbon Concentration Dependence of the Superconducting Transition Temperature and Structure of MgC$_x$Ni$_3$

T. G. Amos[a], Q. Huang[a,b], J. W. Lynn[a], T. He[c] and R. J. Cava[c,*]

[a] *NIST Center for Neutron Research, Gaithersburg, MD 20899, USA*

[b] *Department of Materials and Nuclear Engineering, University of Maryland, College Park, MD 20742, USA*

[c] *Department of Chemistry and Princeton Materials Institute Princeton University, Princeton NJ 08544, USA*

## Abstract

The crystal structure of the superconductor MgC$_x$Ni$_3$ is reported as a function of carbon concentration determined by powder neutron diffraction. The single-phase perovskite structure was found in only a narrow range of carbon content, $0.88 < x < 1.0$. The superconducting transition temperature was found to decrease systematically with decreasing carbon concentration. The introduction of carbon vacancies has a significant effect on the positions of the Ni atoms. No evidence for long range magnetic ordering was seen by neutron diffraction for carbon stoichiometries within the perovskite phase stability range.

* Corresponding author. Tel.: +1-609-258-0016, fax: +1-609-258-6746

E-mail address: rcava@princeton.edu (Robert J. Cava)

**Introduction**

The discovery of superconductivity at 39 K in $MgB_2$ (1) has revitalized interest in intermetallic superconductors. One of the results of research following that discovery was the report of superconductivity near 8 K in $MgCNi_3$ (2). This compound has a perovskite structure analogous to that of $CaTiO_3$. The high Ni content and especially the unusual electronic structure (3), suggest that the superconductivity may occur in close electronic proximity to ferromagnetism. The experimental characterization of $MgCNi_3$ does not yet clearly show the presence or absence of such a relationship (4,5). The superconducting compound can be prepared with variable carbon stoichiometry (2,6,7), providing a potentially good way to probe the properties as a function of electron count. The carbon stoichiometry of the compound actually synthesized is not, however, exactly equal to the nominal starting carbon content of the reactants (2). Therefore a determination of the variation of superconducting properties with carbon content requires determination of the carbon stoichiometry of the actual samples employed in the physical characterization measurements. This stoichiometry is only reliably determined through structural refinement of the perovskite phase. Here we report the crystal structures for $MgC_xNi_3$ in its single-phase stoichiometry range, approximately $0.88 < x < 1$, determined by neutron powder diffraction, and the characterization of the superconducting transition temperature, $T_c$, as a function of C concentration for the same samples. $T_c$ decreases with increasing carbon concentration, but the limit of phase stability is reached before superconductivity fully disappears. The carbon concentration was found to significantly affect the thermal vibration of the Ni atoms, which we attribute to the relaxation of the positions of Ni atoms in the vicinity of carbon vacancies.

**Experimental**

Samples of $MgC_xNi_3$ were prepared as described elsewhere (2). Starting materials were bright Mg flakes, fine Ni powder, and glassy carbon spherical powder. Due to the volatility of Mg encountered during synthesis, 20% excess Mg was employed in the initial mixtures. Nominal compositions of $MgC_xNi_3$ with $x$ = 1.25, 1.05, 1.0, 0.95, 0.9, 0.85 and 0.80 were prepared to span the possible carbon stoichiometry of the perovskite phase. Starting materials were mixed in 3 gram batches and pressed into pellets. The pellets were placed on Ta foil supported in alumina boats, and fired in a quartz tube furnace under a mixed gas of 95% Ar 5% $H_2$. The samples were heated for half an hour at 600°C, followed by one hour at 920°C. After cooling, they were ground with an additional 20% Mg, pressed into pellets, and reheated at 600°C for half an hour and at 920°C for one hour. This was repeated until the neutron powder diffraction measurements showed the perovskite phase to have symmetrical diffraction peaks indicating composition uniformity. This required repeating the above heating procedure between 5 and 10 times.

Superconducting transitions were characterized by zero field cooling in a Quantum Design PPMS magnetometer (8). The superconducting transition temperatures were found by extrapolating the steepest slope of the M vs. T curve to M=0. The uncertainty of $T_c$ was less than 0.05 K. The lowest temperature that can be reached with this instrument is 1.8 K, and therefore superconding transition temperatures below that value could not be measured.

Neutron powder diffraction data were collected using the BT-1 32-counter high resolution diffractometer at the Center for Neutron Research at the National Institute of Standards and Technology. A Cu (311) monochromator, yielding a λ of 1.5402(2) Å was employed. Collimations of 15', 20', and 7' of arc were used before and after the monochromator, and after the sample, respectively. Data were collected at room temperature over a 2-theta range of 3-168°. The crystal structure was refined using the Rietveld method with the General Structure Analysis System (GSAS) software (9). The neutron scattering amplitudes used in the refinements were 0.538, 0.665, and 1.03 ($\times 10^{-12}$cm) for Mg, C and Ni, respectively.

A search for possible magnetic scattering was carried out on the NIST Center for Neutron Research high intensity BT-7 triple-axis spectrometer. The 2θ angular range in these measurements was 3 to 71°, with a fixed incident energy of 13.462 meV and a λ of 2.4649 Å defined by a pyrolytic graphite (PG) monochromator. A PG filter was placed in front of the monochromator to suppress any higher-order wavelength contaminations. Coarse collimations of ~40' were employed to maximize the intensity. Data were collected at temperatures between 1.3 K and room temperature, using 0.2° step sizes. Counting times ranged from 1 to 7 minutes per data point depending on the sample temperature. Each polycrystalline sample was loaded into an aluminum can, mounted in a top loading helium cryostat, with a low temperature capability of 1.3 K. Experimental uncertainties indicated in this work represent one statistical standard deviation, except when noted otherwise.

Figure 1 shows the powder neutron diffraction pattern for one of the single phase compositions, $MgC_{0.9}Ni_3$, in the main panel and one of the carbon compositions below the limit of the perovskite phase stability, $MgC_{0.8}Ni_3$, in the right inset. The latter sample shows, in addition to peaks from the main perovskite superconducting phase, the presence of a second phase with broadened peaks indicating the possibility of compositional variation.

**Results**

MgC$_x$Ni$_3$, as shown in the left inset of Fig. 1, has a perovskite structure with $Pm\overline{3}m$ symmetry (2) with $a \approx 3.81$ Å and atomic positions: Mg (1$a$): 0, 0, 0; C (1$b$): 1/2, 1/2, 1/2; and Ni (3$c$): 0, 1/2, 1/2. The neutron powder diffraction patterns revealed a completely pure single-phase sample for the nominal $x$=0.9 sample. For all other samples, a few additional low intensity peaks were observed and identified to be due to MgO (< 4 weight % of the sample in all cases). The MgO was taken into account as a second phase in the structure refinements in those cases. MgNi$_2$ and Ni impurities were found in the $x$=0.8 sample, which is below the minimum carbon concentration limit of the perovskite phase. The un-reacted C for samples mixed at nominal carbon contents beyond the carbon rich limit of the perovskite phase was present as graphite, evidenced by one broad, weak peak (less than 2% of the (100) perovskite reflection) at 42.5° in the BT-7 data.

The structural refinements showed high correlation between the vibrational parameter $U$ and occupancy parameter $n$ of the C atom (~70%). It was therefore necessary to iteratively refine $n$(C) while keeping $U$(C) fixed, the standard procedure in such cases. The refinements were carried out over a wide range of values for $U$(C) (from 0 to 0.016 Å$^2$). The corresponding values of $n$(C) changed ~ 20 % in this process. A $U$(C) value of 0.006 Å$^2$, found by choosing a value near the lowest $\chi^2$, was fixed, with $n$(C) varied in the final refinements. The refined occupancy parameters of C indicate that the carbon content of the perovskite phase changes from 0.978 to 0.887 as the nominal composition of the starting mixture changes from 1.25 to 0.80. Refinements in which the occupancies of Mg and Ni were varied showed that the site occupancies for these metals were within one standard error of 1, and therefore assumed to be fully occupied in the final calculations. This indicates that there is no Ni or Mg nonstoichiometry or site mixing. The final results are presented in Table I. The observed and calculated neutron diffraction pattern for the $x=0.9$ sample is shown in Fig. 1.

The characterization of the superconducting transitions is shown in Fig. 2. The data for the $x=0.88$ sample are not those of a bulk superconductor. Figure 3 combines the results of the structural refinements and the $T_c$ measurements. It shows the variation of superconducting transition temperature with carbon concentration in the perovskite phase. The figure also depicts the carbon stoichiometry range of the perovskite phase. $T_c$ decreases linearly from 7.3 to 3.4 K as $x$ decreases from 0.98 to 0.91. (The refined carbon content $x$ will be used in the following discussions). The $T_c$ for the ideal compound MgCNi$_3$ is extrapolated to be approximately 8.2 K.

The carbon concentration dependence of the lattice constant $a$, and temperature factors (for Mg and Ni) as a function of $x$ are shown in Fig. 4. The lattice constant decreases linearly from 3.81221(5) Å to 3.79515(5) Å as the refined carbon content $x$ decreases from 0.978 to 0.887. The bottom panel of Fig. 4 shows an increase of $U_{11}$(Ni) with decreasing $x$, which is consistent with an increase in the number of C vacancies. Ni displacements in the plane perpendicular to the Ni-C bonding direction ($U_{22}$ and $U_{33}$) have more freedom than those in the Ni-C bonding direction ($U_{11}$), illustrated in the top panel of Fig. 4. In the presence of the C vacancies, however, the Ni atoms surrounding the vacancy relax their positions, indicated by the significant increase of the displacement factor of $U_{11}$(Ni), and relatively little change for the others. For higher C content samples, the Ni atom has a significantly smaller value for the anisotropic mean square displacement factor in the Ni-C bonding direction ($U_{11}$) compared to those in the plane perpendicular to this bonding direction ($U_{22}$ and $U_{33}$). This behavior is the same as that for the equivalent atoms in many compounds with perovskite structure.

Finally, attempts were made to observe the presence of possible long range magnetic ordering in this system. This was done by looking, as a function of temperature, for the development of new Bragg peaks that might result from antiferromagnetic ordering, or intensity increases in the (100), (110), and (111) nuclear Bragg peaks that might result from ferromagnetic ordering. Diffraction measurements were performed at a series of temperatures in the range of 1.8 K to 300 K, for the samples with $x$=0.968, 0.906 and 0.887. Data were collected using the BT-7 triple-axis spectrometer, optimized for high neutron flux. No evidence was found for any long-range three-dimensional magnetic order, either ferromagnetic or antiferromagnetic in character. We estimate that the detectability limit for a commensurate magnetic structure with our experimental system is less than approximately 0.25 $m_B$/Ni for ferromagnetic ordering and 0.1 $m_B$/Ni for antiferromagnetic ordering.

**Conclusions**

The superconducting transition temperature of the $MgC_xNi_3$ perovskite is approximately linearly dependent on carbon concentration in the relatively narrow stoichiometry range possible for this phase under the synthetic conditions employed here. $T_c$ decreases with increasing carbon deficiency, but the limit of the superconducting perovskite phase stability is reached before superconductivity fully disappears. This suggests that any potential transition from superconductivity to ferromagnetism will be difficult or impossible to reach via variation of the carbon stoichiometry. The clear observation or lack of observation of such a transition is important in clarifying whether or not there is a link between ferromagnetism and superconductivity in this material. However it appears that neither C variation nor Ni site substitution (10) lead to such observations. Mg site substitution, suggested to result in ferromagnetism by theoretical studies (3), has not yet been possible to achieve.

| $x_{nominal}$ | $T_c$ (K) | $x_{neutron}$ | Phase Fraction (%) | | | a (Å) | | $\chi^2$ | $R_p$(%) | $R_{wp}$(%) |
|---|---|---|---|---|---|---|---|---|---|---|
| | | | I | II | MgO | I | II | | | |
| 1.25 | 7.30 | 0.978(8) | 98.0 | 0.0 | 2.0 | 3.81221(5) | - | 1.191 | 4.49 | 5.95 |
| 1.05 | 6.59 | 0.968(4) | 97.3 | 0.0 | 2.7 | 3.81207(2) | - | 1.166 | 3.32 | 3.84 |
| 1.0 | 6.49 | 0.970(5) | 96.8 | 0.0 | 3.2 | 3.81060(2) | - | 1.174 | 3.79 | 4.76 |
| 0.95 | 6.30 | 0.965(5) | 98.5 | 0.0 | 1.5 | 3.80990(2) | - | 1.413 | 4.15 | 5.36 |
| 0.9 | 3.38 | 0.906(5) | 100 | 0.0 | 0.0 | 3.80014(2) | - | 1.612 | 4.28 | 5.53 |
| 0.85 | 3.51 | 0.907(6) | 64.0 | 34.6 | 1.4 | 3.80147(5) | 3.78844(23) | 1.311 | 4.13 | 5.31 |
| 0.8 | <1.8 | 0.887(7) | 56.0 | 37.5 | 3.5 | 3.79515(5) | 3.77802(25) | 1.690 | 3.85 | 4.89 |

**Figure captions**

Fig. 1. Observed (crosses), calculated (solid line) and difference profiles (below) for $MgC_{0.9}Ni_3$. The right inset shows a portion of the observed and calculated profile of the majority phase only for the $MgC_{0.8}Ni_3$ sample (in the inset the lower vertical lines indicate the Bragg positions of the majority phase, the middle lines those of the minority phase and the upper lines those of MgO impurity). The left inset shows the crystal structure of $MgC_xNi_3$.

Fig. 2. Magnetic characterization of the superconducting transitions in $MgC_xNi_3$.

Fig. 3. The dependence of the superconducting transition temperature on carbon stoichiometry in $MgC_xNi_3$.

Fig. 4. Upper panel: Structure of $MgC_xNi_3$ showing the direction of the thermal displacements of the Ni atoms. Figure shows how $U_{11}$(Ni) is affected by the presence and number of C vacancies. Lower panels: The variation of the lattice constant $a$ and the mean square displacement factor $U$ of $MgC_xNi_3$, as a function of the refined carbon content $x$.

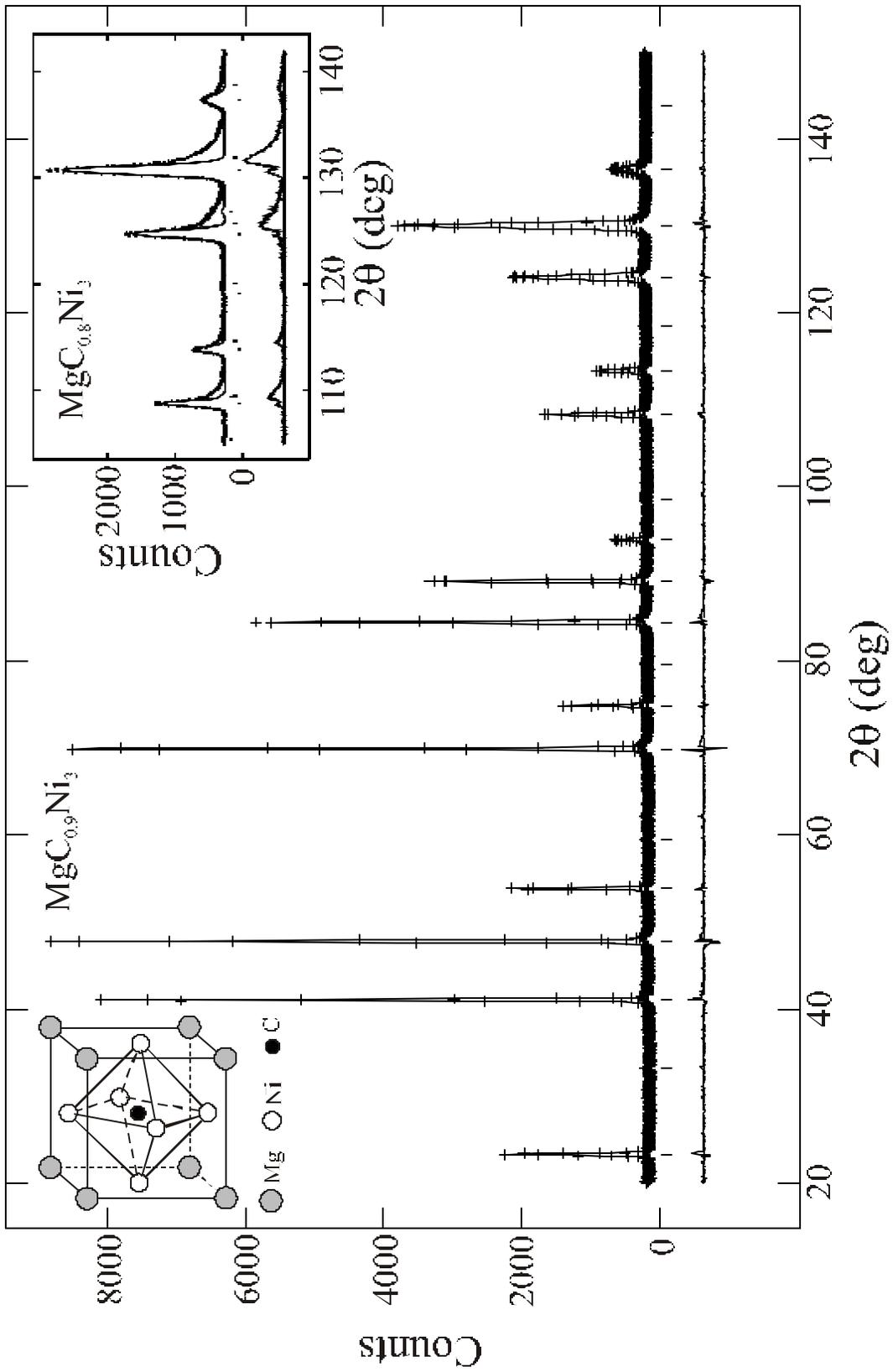

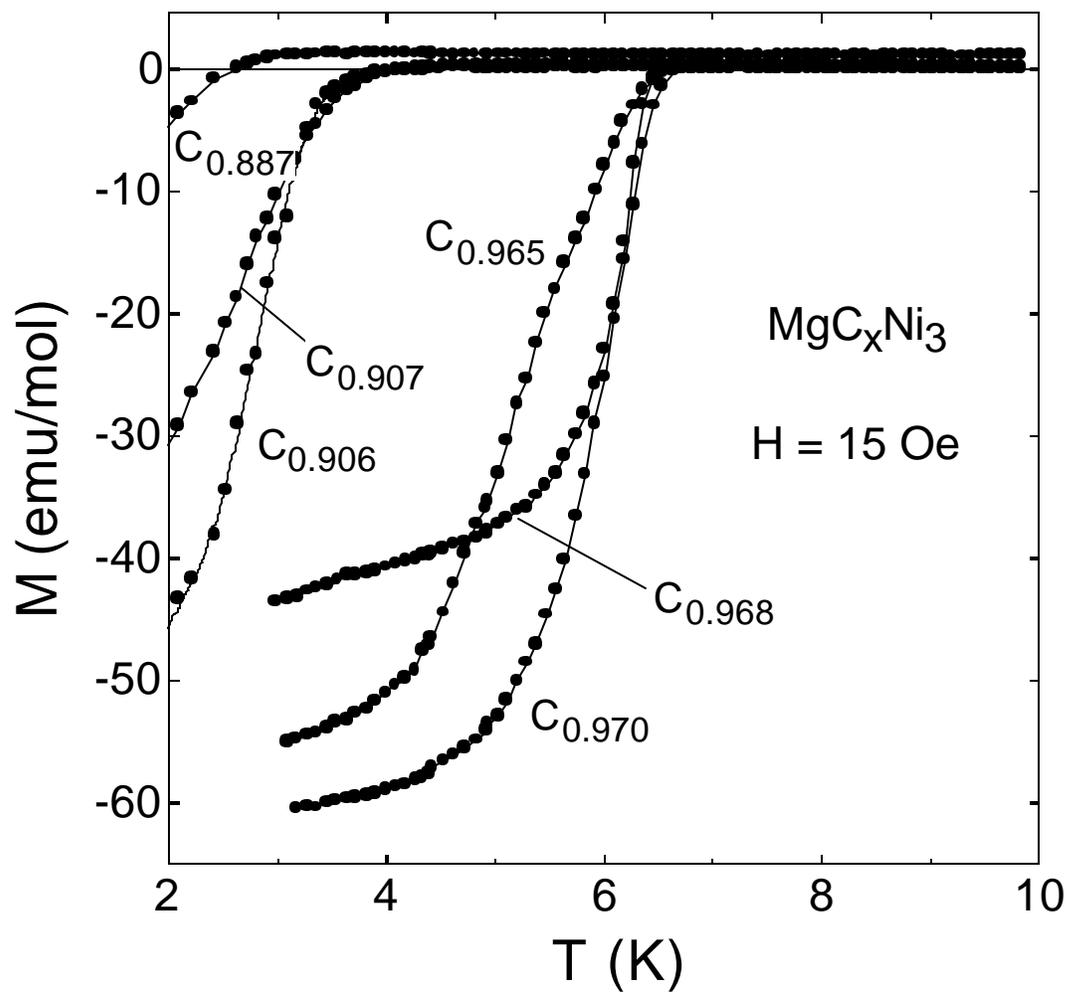

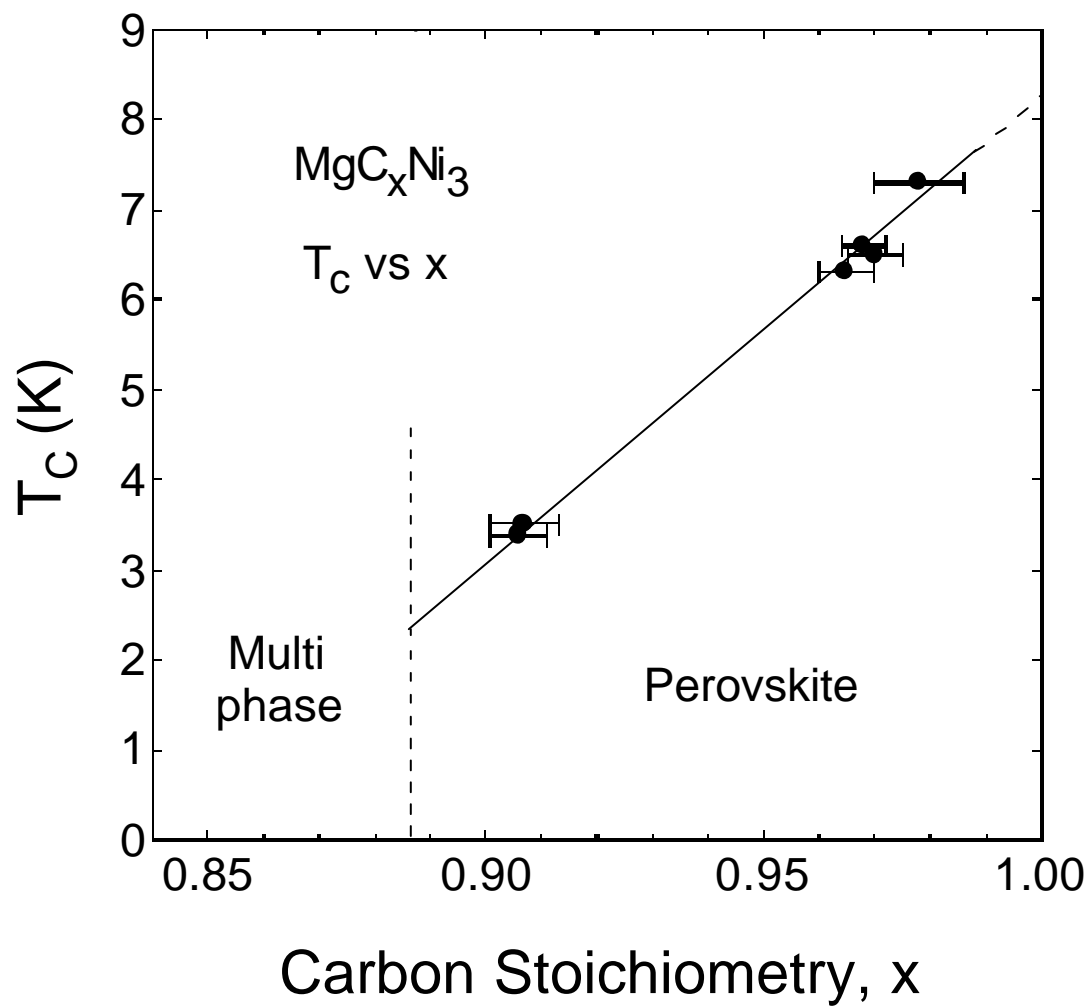

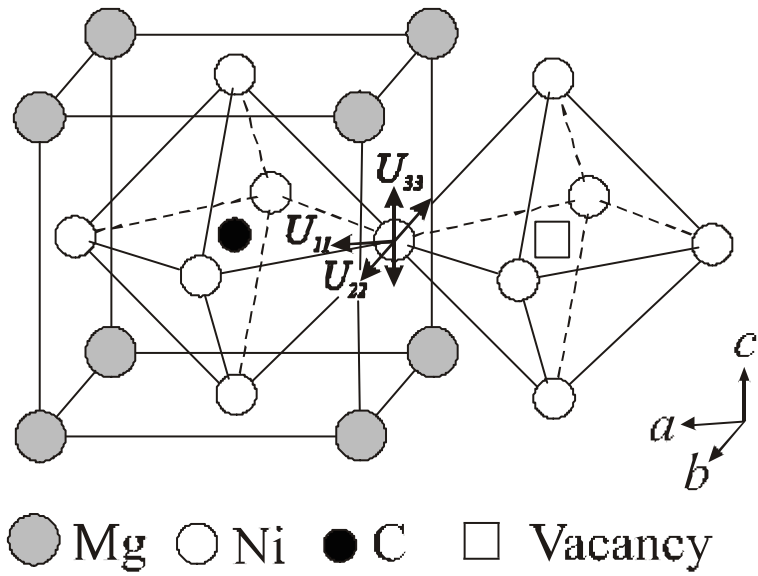

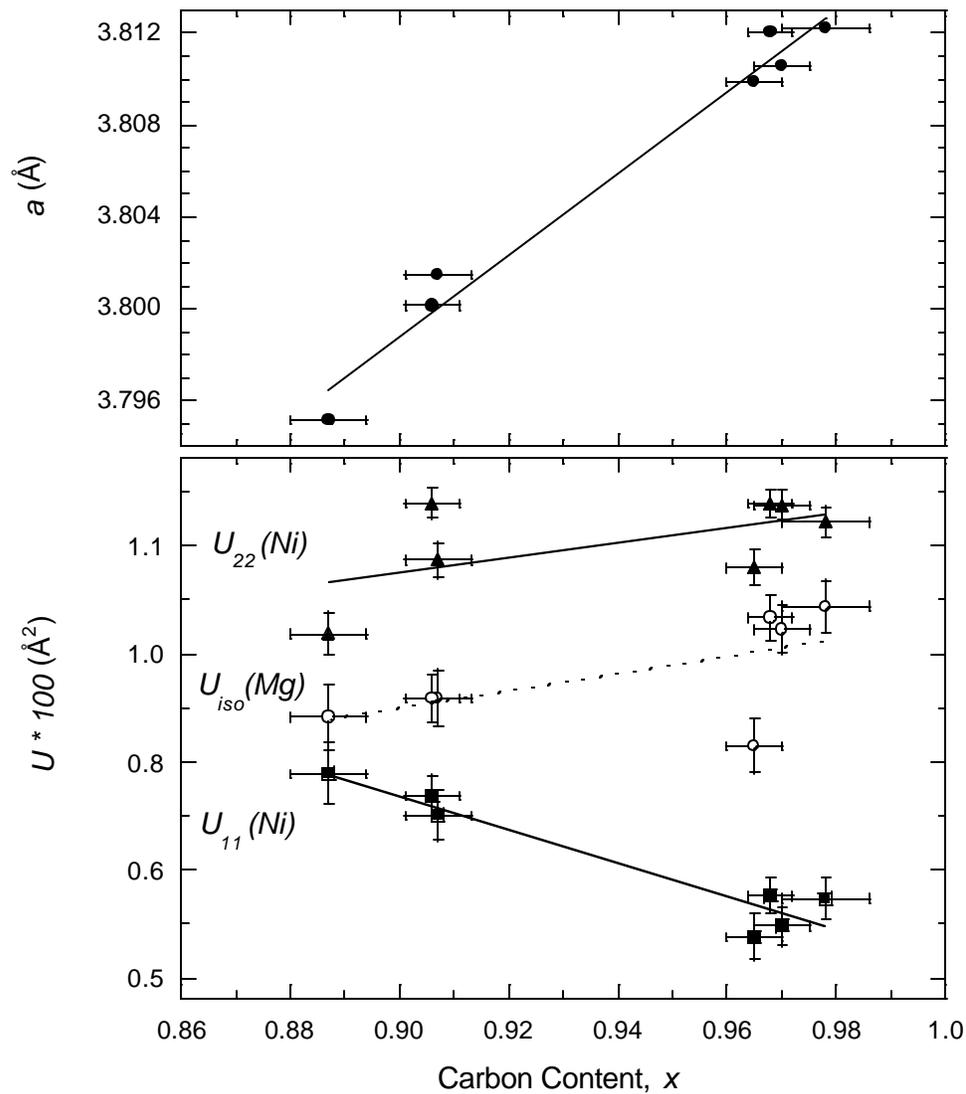